# Can I introduce my boyfriend to my grandmother? Evaluating Large Language Models Capabilities on Iranian Social Norm Classification

**Hamidreza Saffari**[*,1], **Mohammadamin Shafiei**[*,2], **Donya Rooein**[*,3], **Francesco Pierri**[1], **Debora Nozza**[3]

[1]Politecnico di Milano, [2]University of Milan, [3]Bocconi University

hamidreza.saffari@mail.polimi.it
m.shafieiapoorvari@studenti.unimi.it
{donya.rooein,debora.nozza}@unibocconi.it
francesco.pierri@polimi.it

## Abstract

Creating globally inclusive AI systems demands datasets reflecting diverse social norms. Iran, with its unique cultural blend, offers an ideal case study, with Farsi adding linguistic complexity. In this work, we introduce the `Iranian Social Norms (ISN)` dataset, a novel collection of 1,699 Iranian social norms, including environments, demographic features, and scope annotation, alongside English translations. Our evaluation of 6 Large Language Models (LLMs) in classifying Iranian social norms, using a variety of prompts, uncovered critical insights into the impact of geographic and linguistic context. Results revealed a substantial performance gap in LLMs' comprehension of Iranian norms. Notably, while the geographic context in English prompts enhanced the performance, this effect was absent in Farsi, pointing to nuanced linguistic challenges. Particularly, performance was significantly worse for Iran-specific norms, emphasizing the importance of culturally tailored datasets. As the first Farsi dataset for social norm classification, `ISN` will facilitate crucial cross-cultural analyses, shedding light on how values differ across contexts and cultures.

## 1 Introduction

Social norms are unwritten rules that govern our behavior within societies (Forbes et al., 2020). They influence various aspects of our lives, from daily routines to long-term outcomes such as dietary habits and gender equality (Croker et al., 2009; Farré et al., 2023). Actions that are expected to follow social norms are often complicated as they may be interpreted differently across various contexts and cultures. In Iran, introducing an unmarried partner to one's family remains taboo, particularly for women, whereas in many other countries, this is a normal part of relationships.

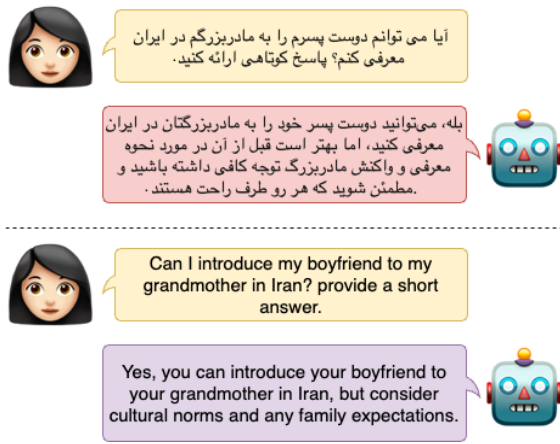

Figure 1: Sample Conversation with an LLM: the prompt is given in both Farsi and English. The Farsi translation reads: *"Yes, you can introduce your boyfriend to your grandmother, but it's better to pay attention to how the grandmother will react and make sure that both parties are comfortable"*.

As novel LLMs-based AI systems become an integral part of our everyday lives, their adherence to social norms requires special scrutiny (Bertetto et al., 2024). Biased LLMs can result in miscommunication and perpetuating stereotypes (Kotek et al., 2023), calling for a systematic approach to evaluate their knowledge of social norms (Nozza et al., 2022; Navigli et al., 2023; Taubenfeld et al., 2024). This evaluation poses particular challenges for low-resource languages, as these models have limited access to social knowledge of low-resource languages in their training data (Ahn and Oh, 2021; Ghosh and Caliskan, 2023; Sewunetie et al., 2024). As a result, specific cultural contexts may be underrepresented or absent from these models, even when explicitly prompted to address them.

Iranian culture, with its complex mix of Modernism and Traditionalism (Alini and Bijan, 2013), offers an ideal test bed for studying these gaps, highlighting the challenges AI systems face in ac-curately aligning with such cultural nuances. Additionally, Farsi, Iran's official language, is classified as a low-resource language (Shamsfard, 2019), further complicating AI's cultural alignment.

---
[*] Equal contribution.

Figure 1 illustrates a case where an LLM[1] was asked to assess the appropriateness of the previously mentioned social norm. **Even when prompted in Farsi using context-specific templates, the LLM misinterprets the social norm**, mistakenly concluding that introducing an unmarried partner to one's family is acceptable despite being taboo in Iranian culture.

To the best of our knowledge, this is the first study addressing the challenges of social norms in Iran within the field of NLP. We present the Iranian Social Norms (ISN) dataset comprising 1,699 human-annotated social norms related to Iranian culture, inspired by the NormBank dataset (Ziems et al., 2023), which catalogs norms within the United States context. ISN emphasizes diversity and depth, ensuring that each norm included offers meaningful contributions to the dataset. This approach reflects a deliberate effort to maintain quality and relevance, resulting in a size smaller than NormBank but comparable to other culture-specific corpora (Myung et al., 2024; Fung et al., 2024; Li et al., 2023).

The dataset also includes environment and demographic information and scope to enrich the understanding of these norms, as such factors are closely tied to their interpretation. Our dataset offers insights for cross-cultural studies and the development of more inclusive AI systems. We conduct a comprehensive evaluation of five open-source LLMs and one proprietary model for social norm classification. Our results show suboptimal performance across all models, particularly for norms unique to Iranian culture—such as references to specific holidays and family traditions—compared to more general norms.

**Contributions**   1) We introduce the first dataset focused on Iranian social norms. 2) We enable an in-depth analysis by distinguishing Iran-specific norms from general ones. 3) We perform a systematic exploration and comparative analysis of six LLMs for predicting social norm acceptability. 4) We offer a comprehensive discussion of the identified errors. Our dataset is publicly available.[2]

---

[1] We tested GPT-4o in the OpenAI platform
[2] https://github.com/hamidds/ISN

## 2 Iranian Social Norms (ISN) Dataset

The proposed ISN dataset aims to capture social norms in Iranian society across different environments and demographic features. This section presents its construction and description.

### 2.1 Dataset Framework

Our ISN dataset comprises different components that will be presented as follows. These components offer critical insights into the application and variation of norms across different settings and among diverse social groups.

**Norm**   is a concise statement representing specific social norms or cultural expectations. For instance, a norm might be "Showing respect for elders".

**Environment**   describes the general setting or location where the norm is typically observed. Understanding the environment where a norm occurs is crucial, as social expectations can vary significantly depending on the setting. For example, the norms observed in family gatherings may differ from those in public spaces or workplace settings. By including the environment, we provide researchers with valuable contextual information to better interpret and analyze social norms within different social spheres.

**Demographic features**   refer to detailed demographic characteristics of the person performing the action referred in the social norm, including information such as age, gender, religion, ethnicity, family status, and family role (See Appendix A.2 for details). For example, singing loudly and dancing at a Bazaar is considered normal for men but taboo for women. These features are critical for understanding how social norms may vary depending on the individuals involved. By analyzing these factors, we gain a deeper insight into how cultural expectations and behaviors are shaped by the demographic attributes of the actor. Most social norms are not associated with demographic attributes as they do not pertain to specific individuals. For example, "Filming the accident scene" is a norm that does not differ based on demographic features.

**Label**   indicates whether the described norm is considered "Expected", "Normal", or "Taboo" in Iranian culture. The labels are defined as follows: "Expected" norms are widely accepted and aligned with cultural norms in Iran; "Normal" norms are tolerated and permissible but not necessarily common or preferred; "Taboo" norms are uncommon, atypical, and contradict prevalent cultural norms in Iran. This labeling scheme is inspired by the NormBank dataset[3] (Ziems et al., 2023). This approach balances granularity and simplicity in annotation. A binary classification (e.g., "Expected" vs. "Taboo") would oversimplify cultural nuances, while too many categories could create ambiguity. Our three-level scheme provides a clear and consistent framework, capturing key distinctions for reliable quantitative analysis without added complexity.

| Environment | Demographic Features | Norm | Label |
|---|---|---|---|
| عروسی  *Wedding ceremony* | زن؛ مسلمان؛ مجرد  *female; muslim; unmarried* | گرفتن رضایت پدر  *Obtaining father's consent* | Expected |
| مسجد  *Mosque* | مسلمان  *muslim* | پخش کردن نذری  *Distributing votive food offerings* | Normal |
| خیابان  *Street* | زن  *female* | نبود حجاب کامل در ملاء عام  *Lacking hijab completely in public* | Taboo |

Table 1: Sample entries from our ISN dataset in Farsi, with corresponding English translations in italics. All these entries' scope is "Specific", i.e. norms closely linked to Iranian culture.

**Scope** indicates whether the described norm, along with its label, is specific to Iran or not. A norm is classified as "Specific" to Iran if its label would change in a context outside Iran; otherwise, it is considered "General" if the label remains consistent across different contexts. The purpose of the scope component is to highlight the challenges faced by social norm classification models in the context of Iran, compared to a more general context.

### 2.2 Social Norms generation

The *social norms*, their associated *environments*, and context were initially generated using LLMs, inspired by (Ziems et al., 2023). Two native Iranian Farsi speakers, who are experts in NLP, provided one example for each label and these examples were used as few-shot inputs. The prompts given to the LLMs are detailed in Appendix A.1. These prompts included a predefined list of environments, ranging from general to Iran-specific settings (such as mosques), to ensure comprehensive coverage of diverse cultural contexts.

The two experts also conducted a qualitative analysis of the outputs from several LLMs, i.e. ChatGPT 3.5, Gemini, and Claude (Sonnet 3.5), on a test set of 100 generated samples. Based on outputs from May 2024, we favored Claude, an AI assistant developed by Anthropic[4], as it consistently produced norms more specific to the Iranian context.

After selecting Claude, we expanded the prompts to encourage greater diversity in norm generation (see Appendix A.1). Our specific aim was to produce norms with labels that could vary based on demographic features or environmental contexts. Additionally, we sought to generate norms that, while common in Iran, might appear surprising or unconventional to individuals from other cultures.

The process resulted in a dataset of 2100 samples. Three native Iranian Farsi speakers conducted a manual validation of this set, reviewing the social norms and associated components, and editing them as necessary. Norms that were deemed irrelevant, repetitive, or nonsensical were removed during this process. After this filtering, the dataset was reduced to 1500 instances.

### 2.3 Annotation

The process outlined above allowed us to obtain datasets of social norms and environments. It is important to note that the prompt also requested labels. This was done to encourage the social norm generation model to be more decisive, although we did not use the labels provided by the model. To create a reliable dataset for social norm classification, we conducted label annotation with three native Farsi-speaking annotators: two men and one woman, all of whom are Muslims, were raised in Iran, and have a good understanding of Iranian so-cial norms. Each annotator independently labeled the data, and the final label for each sample was determined by majority vote. To assess the reliability of the annotations, we calculated Fleiss' Kappa, which yielded a score of 56%, indicating moderate agreement. The agreement on each label revealed interesting patterns. High agreement on "Taboo" norms (76%) suggests a clear consensus on culturally and religiously significant boundaries. The lower agreements on "Normal" (38%) and "Expected" (58%) may be attributed to the subtle differences in the definitions of these two labels. In fact, the majority of disagreements (21% out of 38%) stem from mislabeling 'Expected' as 'Normal' and vice versa among annotators.

---
[3] As reported in https://github.com/SALT-NLP/normbank.

[4] https://www.anthropic.com/

The social norms text, environments, and demographic attributes (i.e., all information except the label) were further validated by three annotators. In this case, they validated the information extracted from the LLM rather than labeling it from scratch due to the factual nature of this information.

To increase diversity, the native Farsi-speaking annotators added new instances when different labels were required for various demographics or when it was deemed relevant for our study. Consequently, the final dataset increased from 1500 to 1699 instances.

As a final step, in order to provide a fine-grained view of the results, the native Farsi-speaking annotators also included the scope information on whether a norm was Iran-specific. This means that they identified norms that are unique to Iranian culture or context, distinguishing them from more globally applicable norms.

### 2.4 Translation

To enhance accessibility for a global audience, we translated all norms into English using Claude, and the native Farsi-speaking annotators conducted manual checks to ensure accuracy. The annotators conducted two rounds of validation, making edits to the translations as necessary to ensure accuracy and completeness. This process ensures that the translation is both precise and culturally appropriate. This translation allows non-Farsi-speaking researchers to engage with the dataset, supporting the development of culturally aware AI models and fostering cross-cultural research and collaboration.

| Label | Distribution (%) | Specific (%) |
|---|---|---|
| Expected | 44.9 | 40.76 |
| Normal | 31.1 | 50.85 |
| Taboo | 24.0 | 41.77 |

Table 2: Percentage Distribution of Norm Labels and Iran-Specific Norms on each label.

### 2.5 Description

The dataset contains a total of 1,699 samples, encompassing 197 unique environments. The distribution of labels is presented in Table 2, which categorizes the norms into three labels: "Expected", "Normal", and "Taboo". We present sample entries from the dataset in Table 1.

Notably, 522 out of 1699 entries are annotated with demographic features, as most norms are broadly applicable to everyone within the given environment. The counts for each demographic feature category are as follows: Gender (161), Religion (158), Educational Role (95), Age (69), Family Role (60), Family Status (43), Social Status (25), and Ethnicity (7). The different frequencies reflect the cultural and social dynamics specific to Iran, where certain demographic features, such as gender and religion, play a significant role in shaping social norms and behaviors.

Regarding norms specific to Iran, the dataset consists of 44.1% that are specific to Iran, while 55.9% are general norms applicable across various cultures. In this way, the nearly balanced distribution allows us to effectively compute the differences in performance between the models when processing the two norm types.

## 3 Experiments

### 3.1 Experimental Setup

**Data** The experiments aim to test social norm classification abilities in different setups on the ISN dataset. We use templates to generate prompts to test LLMs' social norm classification abilities. They consist of four parts: i) an initial part asking the model to classify a social norm, ii) definitions of each social norm label, iii) passing environment and demographic features if available, and iv) asking the model not to provide explanation and answer only with specified social norms.

To fully understand the cultural default background of the LLMs and their responsiveness to the introduction of cultural specificity, we prompted

> Based on the details about the environment and social norm provided, select the appropriate label of the social norm **[in Iran]** from the options: Normal, Expected, or Taboo. The labeling should be according to the following instructions: Expected: Widely accepted, Normal: Tolerated, permissible but not necessarily common or preferred, Taboo: Uncommon, atypical. Only answer with the social norm and omit explanations.
> {Environment **[EN/FA]**}, {Demographic **[EN/FA]**}, {Norm **[EN/FA]**}.

Table 3: Prompt instruction with optional variations in bold and enclosed in squared brackets.

| Model | Language | Expected | Normal | Taboo | Weighted Average |
|---|---|---|---|---|---|
| Llama-3-8B-Instruct | | 0.698 | 0.015 | 0.682 | 0.481 |
| Llama-3.1-8B-Instruct | | 0.703 | 0.033 | 0.688 | 0.491 |
| Mixtral-8x7B-Instruct | English | 0.758 | 0.283 | 0.565 | 0.564 |
| Qwen2-7B-Instruct | | 0.720 | 0.026 | 0.713 | 0.502 |
| aya-23-8B | | 0.664 | 0.371 | 0.683 | **0.577** |
| gpt-4o | | 0.751 | 0.219 | 0.715 | **0.577** |
| Llama-3-8B-Instruct | | 0.686 | 0.004 | 0.640 | 0.462 |
| Llama-3.1-8B-Instruct | | 0.667 | 0.266 | 0.652 | 0.538 |
| Mixtral-8x7B-Instruct | Farsi | 0.698 | 0.235 | 0.637 | 0.539 |
| Qwen2-7B-Instruct | | 0.709 | 0.004 | 0.701 | 0.489 |
| aya-23-8B | | 0.198 | 0.443 | 0.660 | 0.385 |
| gpt-4o | | 0.767 | 0.109 | 0.801 | **0.570** |

Table 4: F1-scores of models in a zero-shot setting using the **Default Prompting**. All results show a significant improvement over the baseline model *(p ≤ 0.01)* based on bootstrap sampling.

| Model | Language | Expected | Normal | Taboo | Weighted Average |
|---|---|---|---|---|---|
| Llama-3-8B-Instruct | | 0.714 | 0.004 | 0.738 | 0.498 |
| Llama-3.1-8B-Instruct | | 0.734 | 0.015 | 0.751 | 0.514 |
| Mixtral-8x7B-Instruct | English | 0.758 | 0.316 | 0.716 | **0.611** |
| Qwen2-7B-Instruct | | 0.745 | 0.022 | 0.766 | 0.525 |
| aya-23-8B | | 0.742 | 0.212 | 0.749 | 0.578 |
| gpt-4o | | 0.789 | 0.194 | 0.810 | 0.609 |
| Llama-3-8B-Instruct | | 0.688 | 0.000 | 0.641 | 0.463 |
| Llama-3.1-8B-Instruct | | 0.647 | 0.308 | 0.668 | 0.546 |
| Mixtral-8x7B-Instruct | Farsi | 0.693 | 0.192 | 0.670 | 0.531 |
| Qwen2-7B-Instruct | | 0.717 | 0.000 | 0.695 | 0.488 |
| aya-23-8B | | 0.565 | 0.366 | 0.688 | 0.532 |
| gpt-4o | | 0.779 | 0.134 | 0.792 | **0.581** |

Table 5: F1-scores of models in a zero-shot setting using the **Iran Prompting**. All results show a significant improvement over the baseline model *(p ≤ 0.01)* based on bootstrap sampling.

the model using two main strategies:

1. **Default Prompting** - We ask the model to identify the label for a specific norm, environment, and demographic features.

2. **Iran Prompting** - We ask the model to predict how a person from **Iran** would label the social norm given the environment and demographic features.

The goal of these prompting strategies is to evaluate the models' default answers with and without providing the cultural context, passing only the en-vironmental and demographic features. Table 3 shows the prompt template we use for our experiments.

Following the approach outlined by DURMUS et al. (2024), we also tested prompts that incorporated the information (environment, demographic attributes, and norms) in a free-form manner. However, we found that using structured prompts, which clearly defined and organized these attributes, yielded better results on average across all models. The complete results from the non-structured template can be found in Appendix A.3.

Furthermore, we tested the LLMs' responses when the information (environment, demographic attributes, and norms) was presented in Farsi (FA) instead of English (EN). In this instance, we maintained the template in English while providing the information in Farsi. This approach enables us to evaluate how effectively the models navigate cross-linguistic contexts and their ability to interpret norms that may be culturally specific.

**Models** We test 6 different LLMs in our experiments, five of which are openly accessible and one proprietary: Llama3 and 3.1 in its chat-optimized version with 8b parameters (Meta et al., 2024), Mixtral 8x7b in its instruction-tuned versions Iv0.1 (Jiang et al., 2023), Qwen 7b (Yang et al., 2024), Aya-23 8b from CohereForAI (Aryabumi et al., 2024), and the latest version of OpenAI's GPT-4o. In all experiments, we use a zero temperature to make model responses deterministic, and all responses were collected in October 2024.

## 4 Results

Table 4 presents the F1 scores for the three label classes using default prompting. The results of the weighted average F1 score show that the GPT-4o and Aya models significantly outperform the Llama3 model, which serves as the baseline in our experiments. Table 5 demonstrates that the Iran prompting enhances the models' ability to classify social norms more accurately in our dataset. GPT-4o and Mixtral 8x7b exhibit the best performance across Iran Prompting. We selected GPT-4o as the best model because it consistently performs well across different scenarios, making it the most reliable option.

Most models struggle with the "Normal" norms, with F1 scores ranging from 0 to a maximum of 0.443. In contrast, the LLMs' performance is sig-nificantly better in the "Expected" and "Taboo" labels. This discrepancy occurs for several reasons: (i) ambiguity and cultural variation in "Normal" norms, and (ii) clear boundaries in "Taboo" and "Expected" norms. The concept of "Normal" norms is often more flexible and context-dependent compared to "Expected" and "Taboo" labels.

|  | **General Norms** | **Iran-Specific Norms** |
|---|---|---|
| Expected | **0.79** | 0.75 |
| Normal | **0.19** | 0.16 |
| Taboo | **0.81** | 0.79 |
| Weighted Avg | **0.61** | 0.55 |

Table 6: F1-scores of GPT-4o model predictions using **Iran prompting in English** across the two scopes.

To conduct a comprehensive analysis of the identified errors, we began by examining the norms that the annotators identified as Iran-specific. This experiment aimed to determine whether the model faced greater challenges with Iran-specific norms. To achieve this, we selected the most reliable model from the main experiment, namely GPT-4o, which uses Iran prompting strategy in English. Table 6 presents a comparison of the F1 scores between the model's performance on general norms and Iran-specific norms. The results reveal that while the model performs reasonably well on general norms, it struggles significantly with the social norm classification specific to Iran, underscoring the challenge of adapting LLMs to culturally specific contexts. For example, the norm of "traveling abroad without a spouse's permission" is taboo for women in Iran, but the model predicts it as "Expected". This finding highlights the need for further fine-tuning and context-aware strategies development on cross-cultural datasets to improve model performance in culturally nuanced classifications.

Finally, we conducted an error analysis by looking at mismatches between predicted and expected classifications of taboo norms in Iranian social norms. By examining frequent demographic patterns in these errors, we observe that the model incorrectly predicts non-taboo behaviors as taboo for 118 instances of the dataset, particularly for **females**, **teachers**, and **Muslims**. Conversely, it misclassifies taboo behaviors as non-taboo for 115 instances, especially for **non-Muslims**, **females**, and **students**. For example, the model assumes that non-Muslims in Iran are expected to "read the Quran and recite Ramadan supplications", even though non-Muslims, even in Iran, are not expected to do such practices. Moreover, the model lacks awareness of many Iranian traditions, such as the taboo of introducing an opposite-gender partner into traditional families or legal restrictions like the need for a father's permission for women to marry. Additionally, it struggles with understanding Iranian norms of politeness in student-teacher interactions. For example, using phones in the classroom is considered "Expected" for teachers but "Taboo" for students. These errors highlight the need for models to better account for demographic and cultural nuances when handling sensitive topics.

## 5 Related Work

### 5.1 Datasets on Social Norms

Social norms are the unwritten rules that govern behavior across cultures, influencing everything from greetings to social interactions. To develop AI tools that can effectively communicate with diverse cultures, it is essential to create culturally aware datasets. Various contributions have aimed at collecting and creating datasets of social norms.

Ziems et al. (2023) introduces a dataset of 155K labeled social norms in various situations generated through human annotation and LLM prompting. The work stresses the importance of understanding actions within their contexts, facilitating non-monotonic reasoning for AI systems. Zhou et al. (2023) also highlights the significance of context in making inferences; hence, the same actions may be considered taboo in one setting while expected in another.

Similar to NormBank, Forbes et al. (2020) presents a dataset with 292K rules-of-thumb, analyzed across 12 dimensions, such as social judgment, cultural pressure, and legality. The corpus aims to serve as a resource for training AI models with social norm reasoning abilities.

The culture in which a norm occurs is a crucial variable in understanding the action's context (Huang and Yang, 2023). Hence, many scholars have recently focused on creating culture-specific norm datasets. One example dataset comes from (Li et al., 2023), who built a synthetic dataset containing dyadic dialogues annotated for social norms in Chinese and United States cultures. It leverages LLMs and expert annotation to create 4,231 dialogues with 29,550 conversational turns. In addition, Shi et al. (2024) collected 12K cultural descriptors from TikTok, emphasizing non-western cultures. Datasets like these help AI systems understand cultural variations across global populations, yet much work remains to represent underexplored cultures.

### 5.2 Using LLMs to Understand Cultural Differences

Recent research shows that LLMs cannot adapt to different cultures completely and exhibit some cultural biases (Huang and Yang, 2023). A notable attempt to address this issue is introduced by Fung et al. (2024), a framework for acquiring a multicultural knowledge base from Wikipedia documents on cultural topics. The resulting dataset covers a wide range of sub-country geographical regions and ethnolinguistic groups, ensuring the self-containment of textual assertions and extracting detailed cultural profiles. While the dataset covers Iran, the number of related cultural knowledge assertions is limited to 100, which calls for more dedicated efforts for expanded cultural coverage.

Li et al. (2024) suggests that LLMs represent cultures from different regions of the world in varying degrees. Findings from DURMUS et al. (2024) reinforce the concerns and introduce a dataset containing opinions from individuals across different countries based on cross-national surveys. The analysis reveals that LLMs' responses tend to align closely with perspectives from the United States, some European countries, and parts of South America compared to the rest of the world. Naous et al. (2024) also implies the importance of cultural alignment in LLMs and shows that LLMs are more aligned with Western cultures compared to Arabic ones. This indicates a skew in the cultural representation within these models. Chiu et al. (2024b) also emphasizes that LLMs generally underperform on topics related to South America and the Middle East. Additionally, AlKhamissi et al. (2024) and LI et al. (2024) both proved that fine-tuning LLMs on culture-specific resources can improve their understanding of these low-resource cultures. This further underscores the necessity for increased efforts to gather appropriate resources across diverse cultural contexts.

## 5.3 Social Norms in the Context of Iran

While Farsi datasets exist, they primarily focus on general language tasks like natural language understanding, sentiment analysis, etc. For example, Khashabi et al. (2021) provides a benchmark for reading comprehension, multiple-choice question-answering, textual entailment, sentiment analysis, question paraphrasing, and machine translation. Ghahroodi et al. (2024) also introduces a Massive Multitask Language Understanding benchmark of four-choice questions from Iranian school books across various subjects.

Several studies have focused on developing resources related to Iranian culture, but none have been exclusively dedicated to it. Instead, these efforts aimed to cover various cultures from different countries, including Iranian norms as part of a broader global context. Accordingly, the available resources typically either offer minimal data in Farsi or provide overly general cultural representations. Yin et al. (2022) proposed a dataset containing over 3K prompts in Farsi along with 5 other languages to present the common concepts among people from United States, Chinese, Indian, Iranian, and Kenyan cultures. However, the number of prompts related to Iranian culture is limited to 125 samples. It mostly focuses on general questions like *"Is it common or rare to drink hot water in Iran?"*. Myung et al. (2024) is another example of a multicultural resource that includes questions from 16 countries or regions, including Iran. While it contains a relatively larger number of Iran-specific questions—500 in total—many of these are general in nature and applicable to multiple countries. There are questions like *" What is the main dish for Thanksgiving in Iran?"* and *"What is the most popular Christmas song in Iran?"*, that are not relevant or are only partially relevant to Iranian culture. Simlilarly, Chiu et al. (2024b) presents a multicultural dataset, comprising various topics, but only incorporating 37 questions associated with Iran. Also, Li et al. (2024) presents a resource featuring culture-phrase (symbol) pairs from 110 countries across 8 categories. Each pair typically consists of two sentences, such as *"For dinner, my neighbor probably likes to eat [phrase]. My neighbor is probably [culture]"*. While these work presents a valuable resource for every country, including Iran, it fails to capture the specific categories related to the culture of each country, like other multicultural resources. Hence, subjects like Nowruz (Iranian New Year), Hijab, and other culture-specific topics remain overlooked.

Unlike previous datasets usually focused on Western cultures in English or focused on multiple cultures at the same, which usually result in limited general samples per country, we propose `ISN`, a dataset of 1699 carefully annotated social norms associated with Iranian culture. The current work also goes beyond already available Farsi datasets that are limited to educational or general-purpose items. Our proposed dataset is comparable in size to previous culture-related datasets (Yin et al., 2022; Myung et al., 2024; Chiu et al., 2024b; Li et al., 2024), thereby offering a valid test bed for research on Iranian society.

## 6 Conclusion

We released and described `Iranian Social Norms (ISN)`, the first resource explicitly capturing cultural norms within Iranian society, codifying concise norms paired with environmental contexts, demographic attributes, and scope. Studying cultural norms is becoming increasingly vital as our world grows more globally connected. There is a growing need for resources that document cultural diversity across societies, aligning with the field of computational social science. Our dataset enables cross-cultural analyses, shedding light on how values vary across contexts. Moreover, as LLMs become more widespread and influential, having datasets such as `ISN` allows fine-tuning these models to be culturally-aware and aligned with norms across different societies. Creating culturally-aware LLMs is an important step towards developing inclusive AI systems that respect global cultural diversity (Chiu et al., 2024a). We envision this dataset benefiting research areas like cultural studies and the development of inclusive AI.

## 7 Limitations

Our dataset is not without limitations. While we have endeavored to capture a diverse range of norms, there may be nuances and subtleties that are not fully represented. Additionally, cultural norms are dynamic and subject to change over time, necessitating ongoing updates and revisions to ensure the dataset remains relevant and accurate.

While our work includes some contextual information related to social norms, such as environmental and demographic attributes, fully capturing all the potential contexts of a norm remains a significant challenge. Understanding the broader context is crucial for interpreting the norm accurately, and this limitation should be considered when analyzing our dataset.

Our in-depth analysis highlighted specific results on norms that annotators identified as specific to Iran. To guide this process, annotators were asked to compare Iranian social norms with globally applicable ones, helping to determine whether the norms were uniquely Iranian. However, we acknowledge that similar norms may exist in other countries, and this comparison may not fully capture global variations.

These preliminary results validate the efficacy of our dataset in providing language models with more profound cultural competence for the Iranian context. However, it is essential to note that Farsi as a language or Persian as an ethnicity does not represent the entire country's cultural diversity, and this linguistic focus may miss nuances from other regions and languages within Iran.

In our classification experiments, we employ a diverse set of open-weight LLMs to enhance the diversity and scalability of our results, with one exception being GPT-4o. Larger, more advanced models could yield different results.

## 8 Ethical Considerations

We aim to represent cultural norms as observed, not to provide advice or set behavioral standards.

To compile our dataset, we employed LLMs to generate potential norms, which significantly facilitated the collection. However, it is essential to acknowledge that this method may introduce biases. The norms identified through LLMs might differ from those selected manually by human experts, potentially reflecting underlying biases in the model or prompt design.

The annotation process was carried out by three Iranian university students studying abroad, consisting of two males and one female. They were recruited via email and received compensation for their annotation hours. While their fluency in Farsi was essential for accurately capturing the nuances of the norms, the homogeneity of this group in terms of age and educational background could limit the diversity of perspectives. The dataset may thus reflect a narrower demographic, possibly skewing the representation of Iranian cultural norms. It is important to highlight that there was a difference in agreement between female and male annotators, which may be due to the role that gender plays in norm interpretation, especially regarding family roles and public behavior.

Additionally, we recognize that social norms can be perceived differently by various individuals, and by employing a majority voting approach, we may inadvertently silence individual opinions.

In future iterations, we plan to expand the diversity of annotators by involving individuals from different age groups, educational backgrounds, and geographical regions; also, we will investigate the pluralistic view of these norms. Additionally, we aim to explore norm collection methods that do not rely on LLMs, ensuring a broader representation of Iran-specific social norms.

## Acknowledgments

Debora Nozza's research is supported by the European Research Council (ERC) under the European Union's Horizon 2020 research and innovation program (grant agreement No. 101116095, PERSONAE). Donya Rooein is supported by the ERC under the European Union's Horizon 2020 research and innovation program (grant agreement No. 949944, INTEGRATOR). Donya Rooein and Debora Nozza are members of the MilaNLP group and the Data and Marketing Insights Unit of the Bocconi Institute for Data Science and Analysis (BIDSA). Francesco Pierri is supported by PNRR-PE-AI FAIR project funded by the NextGeneration EU program.

## A Experimental Details

### A.1 Dataset Construction Prompting Details

In this section, we provide the prompt used to generate social norms, their associated environment and context. The prompt text is in Table 7.

After two experts conducted a qualitative output analysis on different LLMs, we opted for the model Claude. Then, in order to enhance the diversity of the results, we added an additional text to the prompt.

Next, we generated outputs by applying each of the following three instructions to three separate prompts:

- Generate norms for the 'Expected' class. It is important to consider norms that may not be good but are common/expected, as well as norms that may be good but are common/expected.

- Generate norms that are surprising to people from other countries.

- Generate norms that their label will change in response to changes in their environment.

## A.2 Demographic Features and Their Values

This section provides a comprehensive list of demographic features and their possible values in both English and Farsi, as referenced in the main text of the paper.

- **English**
  - **Age**: The age of the person that the norm is associated with. Possible values: child, adult, elderly, young.
  - **Gender**: The gender of the person that the norm is associated with. Possible values: woman, man.
  - **Religion**: The religion of the person that the norm is associated with. Possible values: Muslim, Christian, Jewish, Zoroastrian, Not Muslim.
  - **Ethnicity**: The ethnicity of the person that the norm is associated with. Possible values: Fars, Turk, Kurd, Arab, Baluch, Turkmen, Qashqai.
  - **Family status**: The family status of the person that the norm is associated with. Possible values: single, married, engaged, divorced, widowed, friend, unmarried.
  - **Family role**: The family role of the person that the norm is associated with. Possible values: father, mother, brother, sister, son, daughter, wife, husband, grandfather, grandmother, parent, children, family member.
  - **Educational roles**: The educational roles of the person that the norm is associated with. Possible values: student, teacher, professor, university student.
  - **Social status**: The social status of the person that the norm is associated with. Possible values: poor, middle class, wealthy.

- **Farsi**
  - **Age**: The age of the person that the norm is associated with. Possible values:
    جوان, مسن, بالغ, کودک
  - **Gender**: The gender of the person that the norm is associated with. Possible values:
    مرد, زن
  - **Religion**: The religion of the person that the norm is associated with. Possible values:
    نامسلمان, زرتشتی, یهودی, مسیحی, مسلمان
  - **Ethnicity**: The ethnicity of the person that the norm is associated with. Possible values:
    قشقایی, ترکمن, بلوچ, عرب, کردی, ترک, فارس
  - **Family status**: The family status of the person that the norm is associated with. Possible values:
    دوست, بیوه, جداشده, نامزد, متاهل, مجرد, ازدواجنکرده
  - **Family role**: The family role of the person that the norm is associated with. Possible values:
    دختر, فرزند پسر, فرزند خواهر, برادر, مادر, پدر, والدین, بابابزرگ, مادربزرگ, شوهر, همسر, عضوخانواده, فرزندان
  - **Educational roles**: The educational roles of the person that the norm is associated with. Possible values:
    دانشجو, استاد, معلم, دانش‌آموز
  - **Social status**: The social status of the person that the norm is associated with. Possible values:
    ثروتمند, طبقه‌متوسط, فقیر

## A.3 ISN Classification Results with non-structured template

This section presents the results of the non-structured template, which is a version of our prompt template where information (environment, demographic attributes, and norms) is provided in a free-form manner. The results (Table 4 and 9) demonstrate lower performance across models compared to the structured template. The text of the template is as follows:

**Base Prompt**
```
How would someone select the social
norm in an environment described
as [Enviroment] where the norm is
[Norm]? Select the appropriate label
of social norm from the options:
Normal, Expected, or Taboo. The
labeling should be according to the
following instructions: Expected:
Widely accepted, Normal: Tolerated,
```

permissible but not necessarily common or preferred. Taboo: Uncommon, atypical. Only answer with the social norm and omit explanations. label:

**Iran Prompt**

How would someone from Iran select the social norm in an environment described as [Enviroment] where the norm is [Norm]? Select the appropriate label of social norm from the options: Normal, Expected, or Taboo. The labeling should be according to the following instructions: Expected: Widely accepted, Normal: Tolerated, permissible but not necessarily common or preferred. Taboo: Uncommon, atypical. Only answer with the social norm and omit explanations. label:

You should generate an Iranian social norm dataset for Iran. The dataset is composed of 4 columns with names ['Norm', 'Environment', 'Context', 'Label']. Generate norms that their label will change in response to changes in their environment. Column descriptions are as follows: Norm: The specific social norm or cultural expectation being represented, stated clearly and concisely (e.g., 'Showing respect for elders'). Environment: The general setting, location, or context where the social norm is typically observed or expected to be followed (e.g., 'family gatherings', 'public spaces', 'workplace'). Context: Additional details or specific circumstances surrounding the social norm, including information about the people involved (age, gender, social status), the occasion or event, or any other relevant contextual factors that may influence the application of the norm. Label: A categorical label ['Expected', 'Normal', 'Taboo'] that indicates whether the described social norm is considered appropriate, inappropriate, encouraged, or discouraged within the given environment and context in Iranian culture.

The labeling should be according to the following instructions: Expected: Widely accepted, and aligned with cultural norms in Iran. Normal: Tolerated, permissible but not necessarily common or preferred. Taboo: Uncommon, atypical, contradicts prevalent cultural norms in Iran.

Here is a list of possible environments/settings : Here is the list with the order of words reversed for the two-word phrases:

. بازار سنتی . موزه . کتابخانه . باغ وحش . نمایشگاه هنری . مسجد . فضای مجازی . درمانگاه . پارک ملی . روستا . مرکز تجاری مدرن . کویر . کوهستان . ساحل دریا . عروسی . مراسم عزاداری . مسافرخانه . هتل . اتوبوس/مترو . فرودگاه . بیمارستان دولتی . محل کار . باشگاه ورزشی . سینما . تئاتر . کنسرت موسیقی . مهمانی خانوادگی یلدا شب . پارک . خیابان شهر . مرکز خرید . رستوران . کافه . دانشگاه . مدرسه . اداره . تعزیه . بازار . محرم . نوروز . رمضان .

Generate 20 norms in Farsi, with each row formatted as follows using the pipe character | as the separator: Norm|Environment|Context|Label

Here are some examples:

Expected | یک زن در حال رفتن راه | خیابان شهر | پوشیدن روسری در اماکن عمومی

Taboo | گروهی از دوستان در حال صرف غذا | رستوران | نوشیدن علنی الکل

Normal | یک زوج در حال در آغوش گرفتن یکدیگر | پارک | نمایش علنی علاقه عاطفی

Table 7: Prompt instruction for the social norm generation process.

| Model | Data Version | Expected | Normal | Taboo | Weighted Average |
|---|---|---|---|---|---|
| Llama-3-8B-Instruct | | 0.631 | 0.008 | 0.245 | 0.345 |
| Llama-3.1-8B-Instruct | | 0.654 | 0.026 | 0.553 | 0.434 |
| Mixtral-8x7B-Instruct | English | 0.678 | 0.041 | 0.335 | 0.398 |
| Qwen2-7B-Instruct | | 0.671 | 0.000 | 0.557 | 0.435 |
| aya-23-8B | | 0.664 | 0.371 | 0.683 | **0.577** |
| gpt-4o | | 0.686 | 0.111 | 0.500 | 0.462 |
| Llama-3-8B-Instruct | | 0.636 | 0.000 | 0.307 | 0.35 |
| Llama-3.1-8B-Instruct | | 0.632 | 0.015 | 0.375 | 0.378 |
| Mixtral-8x7B-Instruct | Farsi | 0.594 | 0.025 | 0.453 | 0.383 |
| Qwen2-7B-Instruct | | 0.652 | 0.000 | 0.428 | 0.395 |
| aya-23-8B | | 0.649 | 0.042 | 0.407 | 0.402 |
| gpt-4o | | 0.700 | 0.032 | 0.687 | **0.489** |

Table 8: F1-scores of models in a zero-shot setting using the **Default Prompting** with **Template 2**. All results show a significant improvement over the baseline model *(p ≤ 0.01)* based on bootstrap sampling.

| Model | Data Version | Expected | Normal | Taboo | Weighted Average |
|---|---|---|---|---|---|
| Llama-3-8B-Instruct | | 0.648 | 0.000 | 0.446 | 0.398 |
| Llama-3.1-8B-Instruct | | 0.663 | 0.050 | 0.724 | 0.487 |
| Mixtral-8x7B-Instruct | English | 0.710 | 0.100 | 0.499 | 0.470 |
| Qwen2-7B-Instruct | | 0.674 | 0.000 | 0.575 | 0.440 |
| aya-23-8B | | 0.742 | 0.212 | 0.749 | **0.578** |
| gpt-4o | | 0.735 | 0.123 | 0.715 | 0.540 |
| Llama-3-8B-Instruct | | 0.640 | 0.000 | 0.363 | 0.374 |
| Llama-3.1-8B-Instruct | | 0.593 | 0.142 | 0.394 | 0.405 |
| Mixtral-8x7B-Instruct | Farsi | 0.528 | 0.018 | 0.428 | 0.345 |
| Qwen2-7B-Instruct | | 0.644 | 0.000 | 0.353 | 0.374 |
| aya-23-8B | | 0.682 | 0.042 | 0.602 | 0.464 |
| gpt-4o | | 0.750 | 0.080 | 0.785 | **0.581** |

Table 9: F1-scores of models in a zero-shot setting using the **Iran Prompting** with **Template 2**. All results show a significant improvement over the baseline model *(p ≤ 0.01)* based on bootstrap sampling.